\documentclass[10pt,letterpaper,conference]{IEEEtran}
\usepackage[latin9]{inputenc}
\usepackage{amsmath}
\usepackage{amssymb}
\usepackage{graphicx}

\makeatletter

\usepackage{pstricks, pst-node, pst-plot, pst-circ}
\usepackage{moredefs}
\usepackage{url}

\usepackage{psfrag}
\usepackage{multirow}

\usepackage{amsmath,graphicx}

\usepackage{times}\usepackage{epsfig}
\newcommand{\spaceBelowTab}{\vspace{-4mm}}
\newcommand{\spaceBelowFig}{\vspace{-4mm}}
\newcommand{\spaceBeforeLabel}{\vspace{-1mm}}

\newcommand{\HP}{\text{H\hspace{-0.25mm}P}}
\newcommand{\LP}{\text{L\hspace{-0.25mm}P}}
\newcommand{\sigf}{f}
\newcommand{\fig}{Fig.}
\newcommand{\tab}{Tab.}

\newcommand{\blockbased}{block-based}
\newcommand{\meshbased}{mesh-based}
\newcommand{\Meshbased}{Mesh-based}

\makeatother

\begin{document}
\author{%
{Wolfgang Schnurrer, Thomas Richter, Jürgen Seiler, and André Kaup}%
\vspace{1.6mm}\\
\fontsize{10}{10}\selectfont\itshape 
Multimedia Communications and Signal Processing, University of Erlangen-Nuremberg\\ 
Cauerstr. 7, 91058 Erlangen, Germany\\
\fontsize{9}{9}\selectfont\ttfamily\upshape 
%
\{schnurrer, richter, seiler, kaup\}@lnt.de
%
}

\title{Analysis of Mesh-Based Motion Compensation in  Wavelet Lifting of
Dynamical 3-D+t CT Data}

\maketitle

\begin{abstract}
Factorized in the lifting structure, the wavelet transform can easily
be extended by arbitrary compensation methods. Thereby, the transform
can be adapted to displacements in the signal without losing the ability
of perfect reconstruction. This leads to an improvement of scalability.
In temporal direction of dynamic medical 3-D+t volumes from Computed
Tomography, displacement is mainly given by expansion and compression
of tissue. We show that these smooth movements can be well compensated
with a mesh-based method. We compare the properties of triangle
and quadrilateral meshes.  We also show that with a mesh-based compensation
approach coding results are comparable to the common slice wise coding
with JPEG~2000 while a scalable representation in temporal direction
can be achieved.
\end{abstract}



\section{Introduction}

In wavelet-based video coding, motion compensation methods are used
for better exploiting the correlation between adjacent frames. For
a compensated transform in temporal direction this is called Motion
Compensated Temporal Filtering (MCTF) while a compensated transform
in view direction is called Disparity Compensated View Filtering (DCVF)
\cite{garbasTCSVT}. The video signal is transformed along the motion
trajectories. While \blockbased{} methods are widely used, \meshbased{}
compensation methods have also been proposed. In \cite{secker2003}
triangle-based mesh compensation is introduced in the lifting structure
and is shown to be superior to \blockbased{} compensation. For wavelet-based
approaches a \meshbased{} compensation comes with the advantage that
it can be inverted without causing unconnected pixels compared to
block-based compensation methods. In \cite{nosratinia1996} a mesh-based
compensation method is proposed in an interframe coding framework
for Magnetic Resonance Image (MRI) data. Even though there is no wavelet
transform in this direction, the residuals are coded with a wavelet-based
coder. Mesh-based methods work well for smooth motion vector fields.

While areas that become occluded can be compensated by shrinking the
size of the corresponding patches, the problems are the dis-occluded
areas. This is especially the case when the dis-occluded areas contain
structure that cannot be predicted from the neighboring patches. In
a dynamic medical Computed Tomography (CT) volume, e.g., a cardiac
volume, for adjacent slices in temporal direction the primary occurring
displacement can be described by expansion and compression of tissue.
We assume that there are neither occlusions nor dis-occlusions as
there are usually no moving objects like in video sequences. An approach
for compensation of smooth deformation based on a potential field
has already been shown to be feasible \cite{weinlich2012}. This kind
of smooth displacement seems to be promising for a mesh-based compensation
method.

\begin{figure}
\includegraphics[width=0.49\columnwidth]{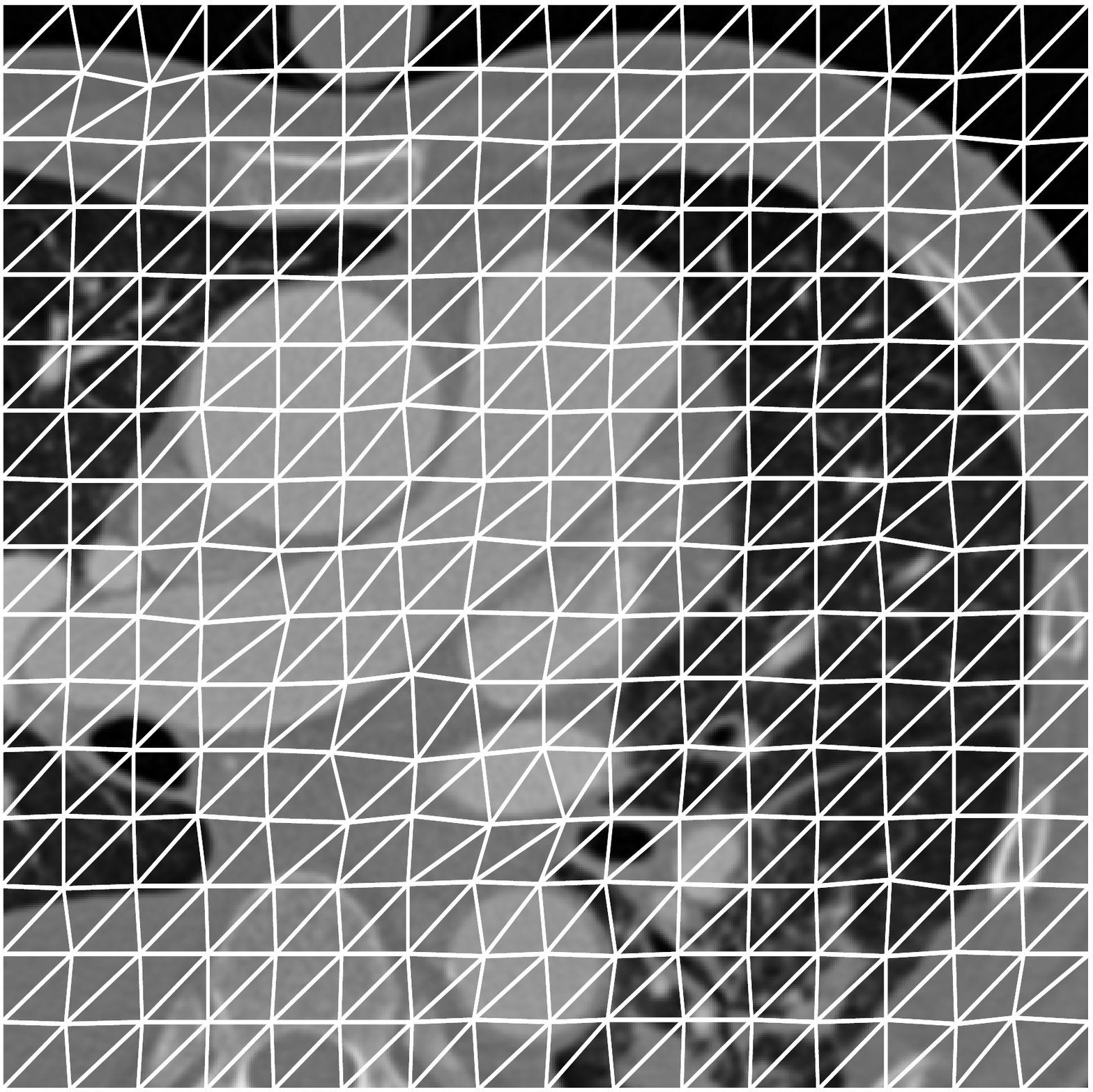}\hfill{}\includegraphics[width=0.49\columnwidth]{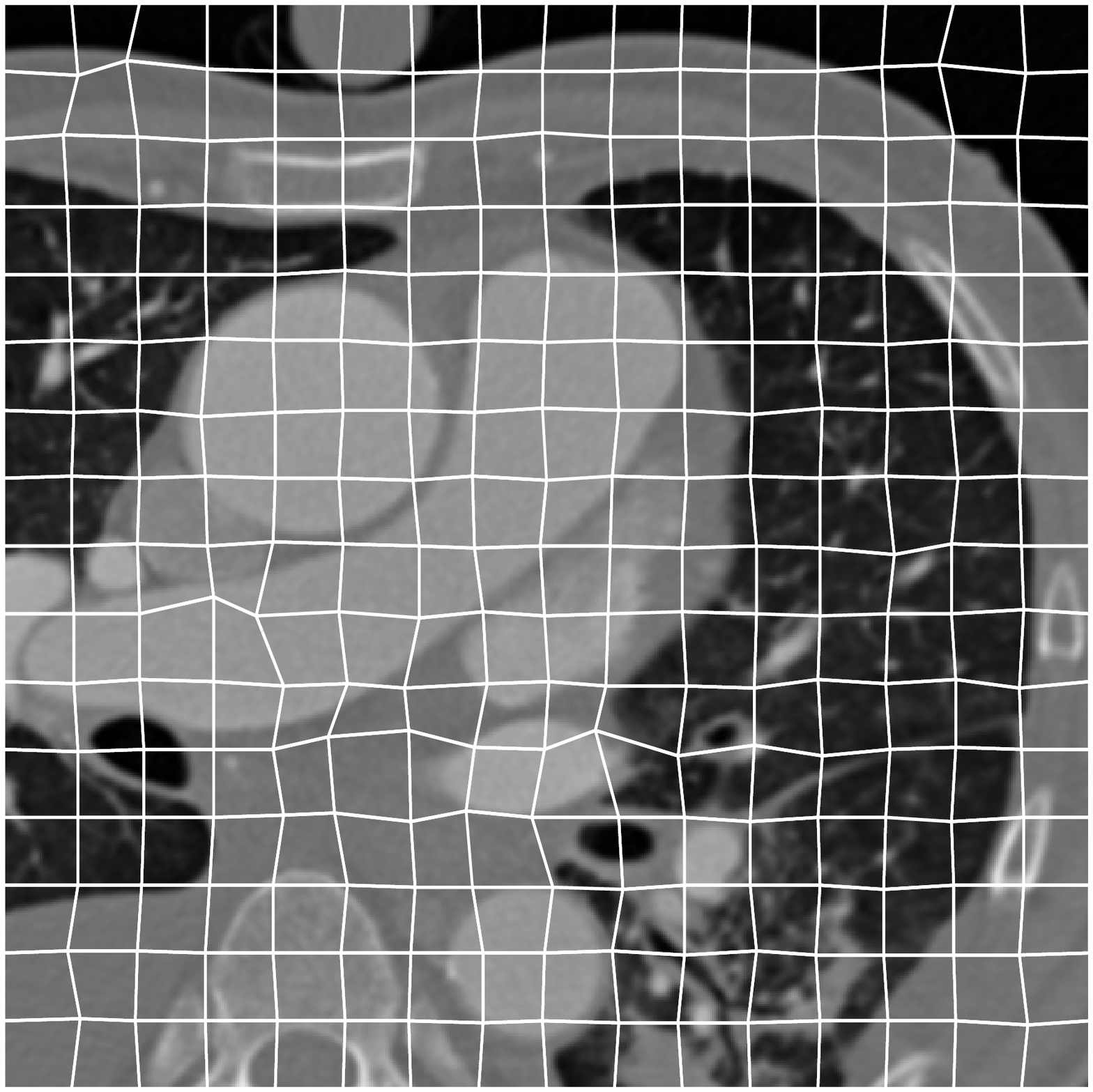}

\spaceBeforeLabel\spaceBeforeLabel\protect\caption{\label{fig:Mesh-based-compensation}Mesh-based compensation, left:
deformed triangle mesh, right: deformed quadrilateral mesh}

\spaceBelowFig
\end{figure}

In \fig{}~\ref{fig:Mesh-based-compensation}, a deformed triangle
mesh is shown as overlay of the reference slice on the left. On the
right, a deformed quadrilateral mesh is shown as overlay on the reference
slice. The reference slice is then warped according to the mesh to
obtain a predictor for the current slice.

Basically, a \meshbased{} approach corresponds to a sampling of the
motion vector field. The vector field is sampled on the grid points
of the mesh and the values in between are obtained by interpolation.
A mesh has two basic parameters that have a significant influence
on its properties. The geometry is one parameter and it defines the
location of each grid point. An active mesh can adapt to the content
of the image. For still image coding one proposed approach is for
example to start with grid points in the image corners and subsequently
add a new grid point where the approximation error can be reduced
to the most extend \cite{adams2008}.  In this paper we do not optimize
the mesh geometry and thus use a regular grid. The other parameter
is the topology. The topology defines the connection of the grid points.
In \cite{bozinovic2004} a mesh topology is proposed that improves
the motion compensation for video sequences. The problem of improper
motion compensation, especially at image boundaries, due to occlusion
and dis-occlusion is addressed. Adjacent slices in a medical volume
are aligned so there is usually no displacement between adjacent slices
that corresponds to horizontal or vertical camera shifts. We will
 compare two different kinds of topology, namely a triangle-based
and a quadrilateral mesh. The topology affects the kind of transformation
that is applied to the patches \cite{nakaya1994}. While triangle-based
methods lead to affine transforms, quadrilateral meshes lead to bilinear
transforms of the patches \cite{sullivan1991}. We will examine which
of these two approaches is able to better compensate the displacement
in the temporal direction of the dynamic medical CT volume.

An important task of \meshbased{} compensation is the estimation
of displacement of the grid points. In contrast to \blockbased{}
methods, the motion vector of a grid point has influence on all its
neighboring patches in the mesh. Thus, the choice of the motion vectors
of the grid points of these neighboring patches is influenced. That
means that for the optimal solution \emph{all} combinations of motion
vectors for all grid points would have to be evaluated. It is assumed
that the global optimum can be sufficiently approximated as it is
not practicable to test all possible combinations. While in \cite{fowler2000qccpack}
a \blockbased{} estimation is used, \cite{sullivan1991} proposes
an iterative estimation and \cite{nakaya1994} combines the two approaches
by refining the \blockbased{} approach.

\Meshbased{} methods have not been analyzed for compensated wavelet
lifting of dynamic 3-D+t CT volumes so far. In this paper we compare
triangle and quadrilateral mesh-based compensation in the lifting
structure with focus on a wavelet transform in temporal direction
of a dynamic medical CT volume of the heart. We further show that
with a \meshbased{} compensation approach coding results comparable
to the common slice wise coding with JPEG~2000 \cite{ITU-T-800}
can be achieved while a scalable representation in temporal direction
can be achieved.

In Section~2 we review the triangular and the quadrilateral mesh-based
compensation and discuss the integration into the wavelet lifting.
The description of our simulation and results are given in Section~3.
Section~4 concludes this study.

\section{Mesh-based Compensation in the Lifting Structure}

\subsection{Lifting Structure of the Haar Wavelet}

\begin{figure}
\psfragscanon

\psfrag{time}{time}
\psfrag{ft1}{$f_{2t}$}
\psfrag{ft2}{$f_{2t+1}$}
\psfrag{P}{$\mathcal{P}$}
\psfrag{U}{$\mathcal{U}$}
\psfrag{HPt}{$\HP_{t}$}
\psfrag{LPt}{$\LP_{t}$}

\includegraphics[width=0.99\columnwidth]{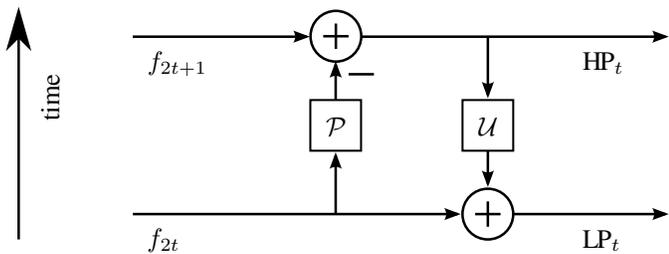}

\psfragscanoff

\spaceBeforeLabel\protect\caption{\label{fig:Lifting-structure}Lifting structure}

\spaceBelowFig
\end{figure}

The lifting implementation of the wavelet transform, shown in \fig{}~\ref{fig:Lifting-structure},
allows the integration of arbitrary compensation methods into the
transform \cite{garbasTCSVT}. The slices of the original volume $\sigf_{i,t}$
are indexed by slice number $i$ in z-direction and time step $t$
in temporal direction. We examine a transform in temporal direction
and thus use a fixed slice number $i$. So for simplicity we denote
the adjacent slices in temporal direction by $\sigf_{t}$ only. The
analysis step consists of a prediction step $\mathcal{P}$ and an
update step $\mathcal{U}$. In the prediction step $\mathcal{P}$,
the highpass coefficients $\HP_{t}$ of the Haar wavelet are computed
to
\begin{equation}
\HP_{t}=\sigf_{2t+1}-\left\lfloor \mathcal{W}_{2t\rightarrow2t+1}\left(\sigf_{2t}\right)\right\rfloor .\label{eq:H-Haar}
\end{equation}
The warping operator $\mathcal{W}_{\text{ref}\rightarrow\text{cur}}$
denotes the computation of a predictor for the current slice with
index `cur' based on the reference slice with index `ref' \cite{garbasTCSVT}.
In the update step $\mathcal{U}$, the lowpass coefficients $\LP_{i}$
are computed to 
\begin{equation}
\LP_{t}=\sigf_{2t}+\left\lfloor \frac{1}{2}\mathcal{W}_{2t+1\rightarrow2t}\left(\HP_{t}\right)\right\rfloor \label{eq:L-Haar}
\end{equation}
using the results from the prediction step. This leads to a significant
reduction of computational complexity. As the index of $\mathcal{W}$$ $
in \eqref{eq:L-Haar} shows, the compensation has to be inverted in
the update step to achieve an equivalent wavelet transform. The introduction
of rounding operators in the lifting structure further allows perfect
reconstruction \cite{calderbank1997}. We use an additional rounding
operator for the computation of the $\HP$-band \eqref{eq:H-Haar}
to avoid fractional values from the warping. Thus, the original volume
can be reconstructed from the wavelet coefficients without loss. This
makes the transform very feasible, e.g., for medical image data.

\subsection{Triangle and Quadrilateral Mesh}

In mesh-based compensation, a mesh is laid over both the reference
slice and the current slice as shown in \fig{}~\ref{fig:Mesh-based-compensation}.
The compensation is computed by the deformation of the image according
to the motion vectors of the grid points. In the prediction step,
the reference image is deformed according to the movement of the grid
points. The result of the warping operator $\mathcal{W}_{2t\rightarrow2t+1}$
is a predictor $p_{2t+1}$ for the current slice $f_{2t+1}$ based
on the reference slice $f_{2t}$ and can be written as
\begin{equation}
p_{2t+1}\left(x,y\right)\hspace{-1mm}=\hspace{-1mm}\mathcal{W}_{2t\rightarrow2t+1}\left(f_{2t}\right)\hspace{-1mm}=\hspace{-1mm}\tilde{f}_{2t}\left(m\left(x,y\right),n\left(x,y\right)\right)\label{eq:warping}
\end{equation}
 where the vector fields $m\left(x,y\right)$ and $n\left(x,y\right)$
describe the deformation of the slice. The tilde indicates the bilinear
interpolation when intensity values are needed on fractional positions
of the reference slice. The interpolation for the values between the
grid points is dependent on the used mesh. We give a brief overview
of the two kinds of meshes that are compared in this paper. In \cite{nakaya1994}
a detailed description of triangle and quadrilateral meshes is given.

A triangle mesh leads to an affine transform. The motion vectors between
the grid points are interpolated between the three grid points of
the current triangle. The resulting interpolated vector field has
a patch-wise constant slope. The corresponding vector fields $m$
and $n$ are computed to 
\begin{align}
m\left(x,y\right) & =a_{i1}x+a_{i2}y+a_{i3}\label{eq:warping-affine}\\
n\left(x,y\right) & =a_{i4}x+a_{i5}y+a_{i6}.\nonumber 
\end{align}
The coefficients $a_{i1}$ to $a_{i6}$ are the entries of the affine
transformation matrix.

A quadrilateral mesh leads to a bilinear transform. The motion vectors
between the grid points are interpolated between the four grid points
of the current quadrilateral mesh. The corresponding vector fields
$m$ and $n$ are computed to
\begin{align}
m\left(x,y\right) & =a_{i1}xy+a_{i2}x+a_{i3}y+a_{i4}\label{eq:warping-bilinear}\\
n\left(x,y\right) & =a_{i5}xy+a_{i6}x+a_{i7}y+a_{i8}.\nonumber 
\end{align}
Again, the coefficients $a_{i1}$ to $a_{i8}$ correspond to the bilinear
transformation matrix.

\subsection{Grid Point Motion Estimation}

\begin{figure}
\includegraphics[width=0.98\columnwidth]{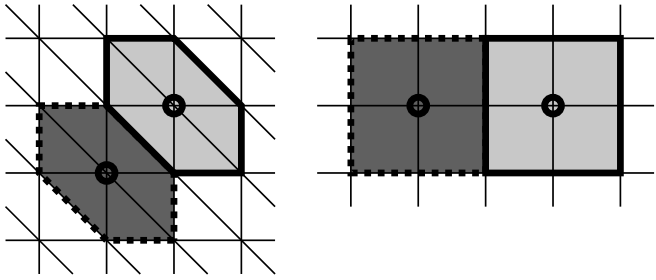}

\spaceBeforeLabel\protect\caption{\label{fig:independentGridPoints}Two grid points with non overlapping
influence area, left: triangle mesh, right: quadrilateral mesh}

\spaceBelowFig
\end{figure}

An important task of \meshbased{} compensation is the estimation
of the displacement of the grid points. For \blockbased{} motion
estimation the optimal motion vector of every block can be searched
independently. In contrast to that, the motion vector of a grid point
has influence on all its neighboring patches in the mesh. This is
shown as gray areas around the marked grid points in \fig{}~\ref{fig:independentGridPoints}.
Thus, also the choice of the motion vectors of the grid points of
these neighboring patches is influenced. That means that for the optimal
solution \emph{all} combinations of motion vectors for all grid points
have to be evaluated. This is computationally by far too complex.
Gradient-based methods were proposed that iteratively refine the motion
vectors of the grid points. These can lead to a suboptimal solution,
because local extrema can avoid the achievement of the global optimum.
In \cite{sullivan1991} an iterative sequential refinement is proposed
for quadrilateral meshes. The vector field is initialized with zero
displacement. Refinement means that the motion vector of a grid point
is modified by one pixel in every direction and the position with
the smallest error metric is chosen. In every iteration all active
grid points are sequentially refined. A grid point is active if either
the grid point itself or one of its neighbors was updated in the previous
iteration. The search terminates either by convergence or when a given
number of iterations is reached.

In \cite{nakaya1994} a parallel approach is proposed for the refinement
procedure. As shown by the gray areas in \fig{}~\ref{fig:independentGridPoints}
the area of influence of one grid point is limited so independent
grid points can be processed in parallel.  To further speed up the
refinement, a coarse estimation is proposed as initialization. For
the coarse estimation a \blockbased{} motion search is performed.
The blocks are centered on each grid point in the current slice and
the initial motion vectors are searched in the reference slice.

\subsection{Inversion for the Update Step}

\begin{figure}
\psfragscanon
\psfrag{wrp}{warping}
\psfrag{inverse}{inverse}

\includegraphics[width=0.98\columnwidth]{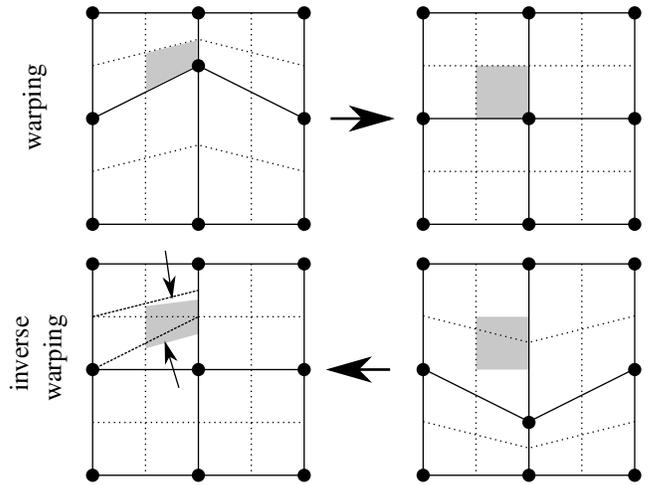}

\psfragscanoff

\spaceBeforeLabel\protect\caption{\label{fig:Approximation-error}Occurring error by approximation of
the inversion is marked by the two arrows in the lower left image}

\spaceBelowFig
\end{figure}

In the update step of the compensated lifting \eqref{eq:L-Haar} the
compensation from the prediction step \eqref{eq:H-Haar} has to be
inverted to obtain an equivalent wavelet transform. The \meshbased{}
approach has the advantage that it is invertible \cite{secker2002}.
For a correct inversion of the \meshbased{} compensation there are
two kinds of interpolation necessary. In the compensation step the
current slice is predicted from the reference slice by evaluating
the reference slice on fractional positions. The corresponding values
are calculated by interpolation from the intensity values available
at the integer positions. In the first step of the inversion, a warping
of the intensity values from integer positions in the current slice
to fractional positions in the reference slice is done. In the second
step, an interpolation from that non regular distribution of sample
points is needed to obtain the intensity values at integer positions
in the reference slice. This interpolation is quite complex.

To avoid this kind of interpolation procedure we use an approximation
from \cite{secker2002} for the inversion of the \meshbased{} compensation
and have to accept the error that is made thereby. Instead of calculating
the inversion of the mesh warping we take the negative values of the
motion vectors at the grid points. This approximation introduces an
error as shown in \fig{}~\ref{fig:Approximation-error}. The smaller
the patches of the mesh and the bigger the deformation, the larger
is the occurring error.

\section{Simulation and Results}

We use a dynamic \textit{cardiac}\footnote{The CT volume data set was kindly provided by Siemens Healthcare.}
3-D+t CT data set. In our simulation we perform one wavelet decomposition
in temporal direction and analyze the performance of the mesh-based
compensation methods with different parameters. We further compare
the \meshbased{} compensation to the \blockbased{} compensation.
The CT data set has a resolution of $512\times512$ in $x$-$y$,
130~slices and 10 time steps and a bit depth of 12~bit per voxel.
We took the 10 temporally adjacent slices at one spatial position
for our simulation to evaluate the behavior of the compensation methods.

For the \meshbased{} compensation method, we use two sets of parameters
with a grid size of 8~pixels and 16~pixels, respectively. For the
\blockbased{} coarse estimation, the blocks centered at the grid
points are set equal to the grid size and a search range of 7~pixels
respectively 9~pixels is used. For the iterative refinement we use
a step size of one pixel in each direction.   For comparison we
also computed a \blockbased{} compensation with a block size equal
to the grid size. The search range is set equal to the coarse estimation.
For the inversion of the \blockbased{} method, we use a nearest neighbor
interpolation of the motion vector field \cite{bozinovic2005} to
fill the unconnected pixels.

With a smaller grid size, the quality of the predictor can be improved
in the compensation step but due to the approximation of the inversion
step the error stays the same. So we show the results for a grid size
of 16~pixels.

\begin{figure}
\includegraphics[width=0.49\columnwidth]{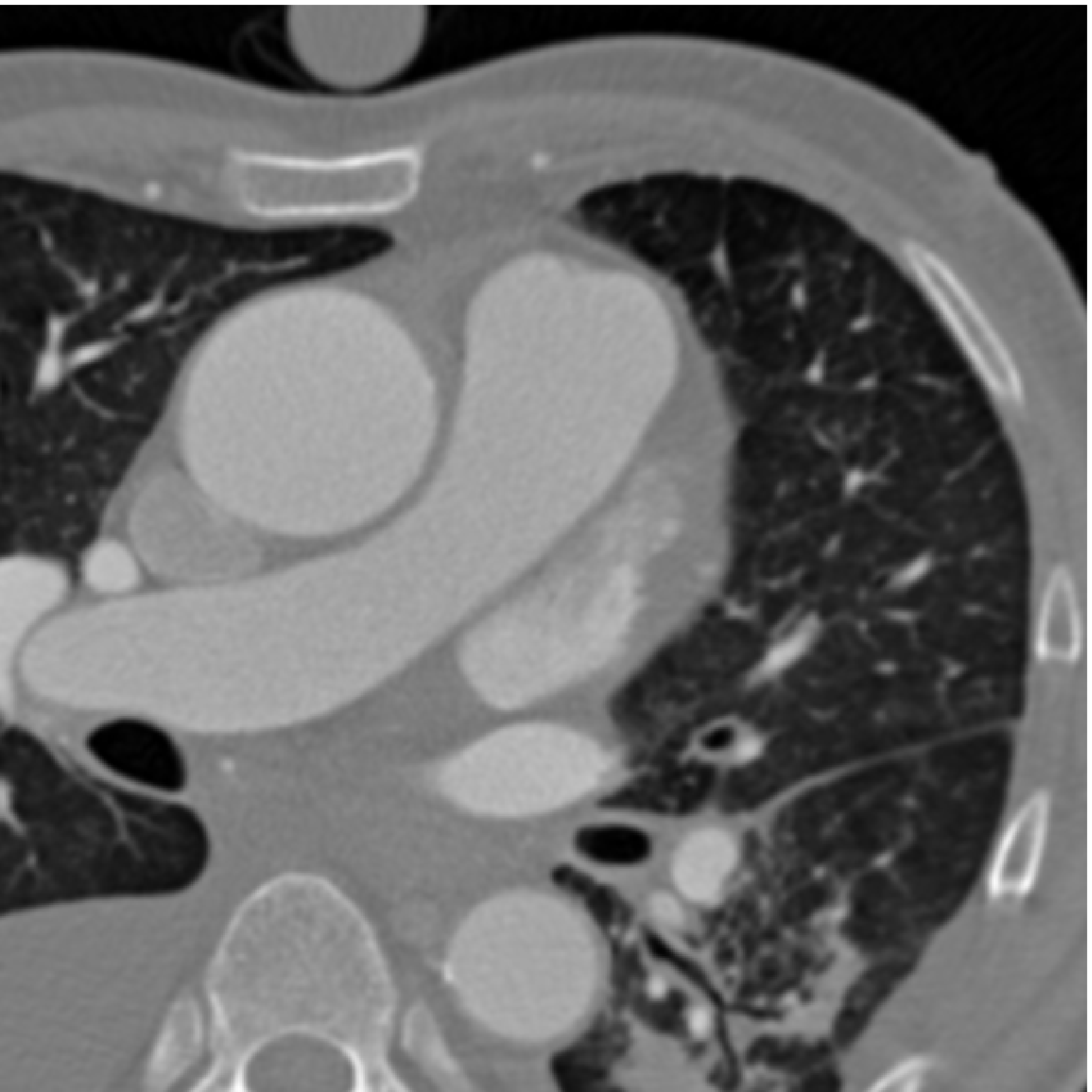}\hfill{}\includegraphics[width=0.49\columnwidth]{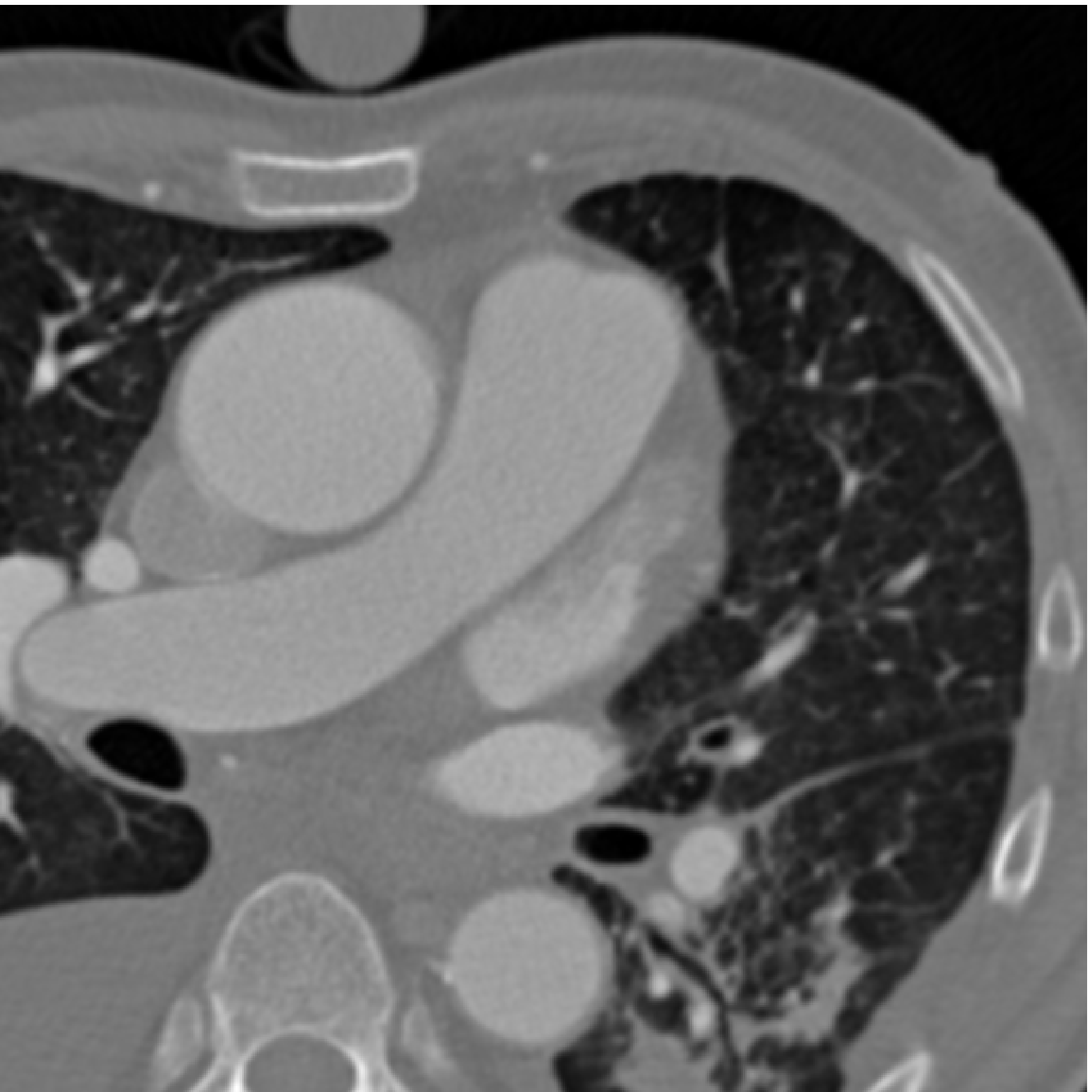}

\hfill{}(a) reference slice $f_{2t}$\hfill{}(b) current slice $f_{2t+1}$\hfill{}

\spaceBeforeLabel\protect\caption{\label{fig:simulation-example}Two adjacent slices in temporal direction}

\spaceBelowFig
\end{figure}

We show the performance of the methods on the two adjacent slices
in temporal direction that are shown in \fig{}~\ref{fig:simulation-example}.

\begin{figure*}
\includegraphics[width=0.32\textwidth]{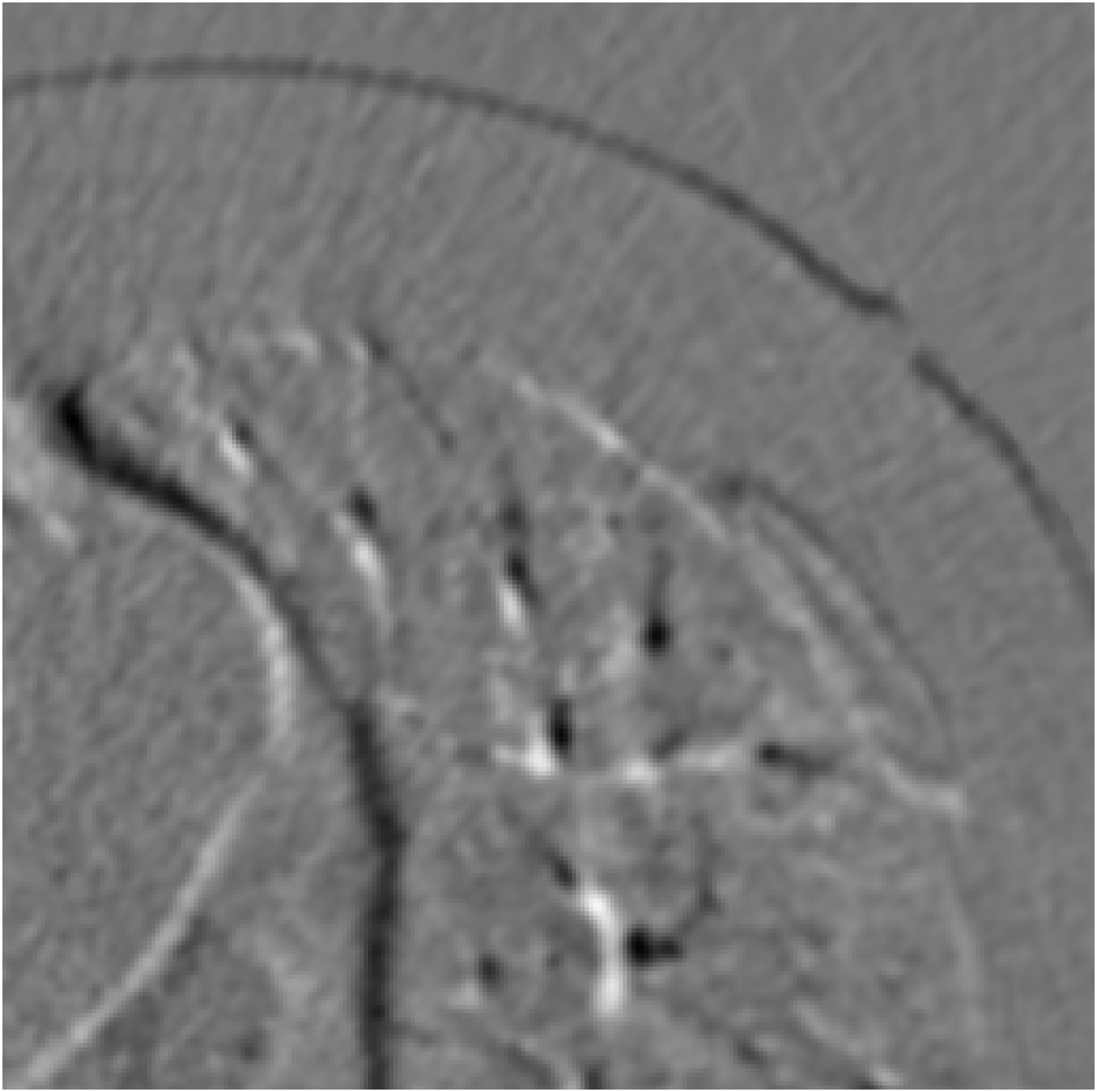}\hfill{}\includegraphics[width=0.32\textwidth]{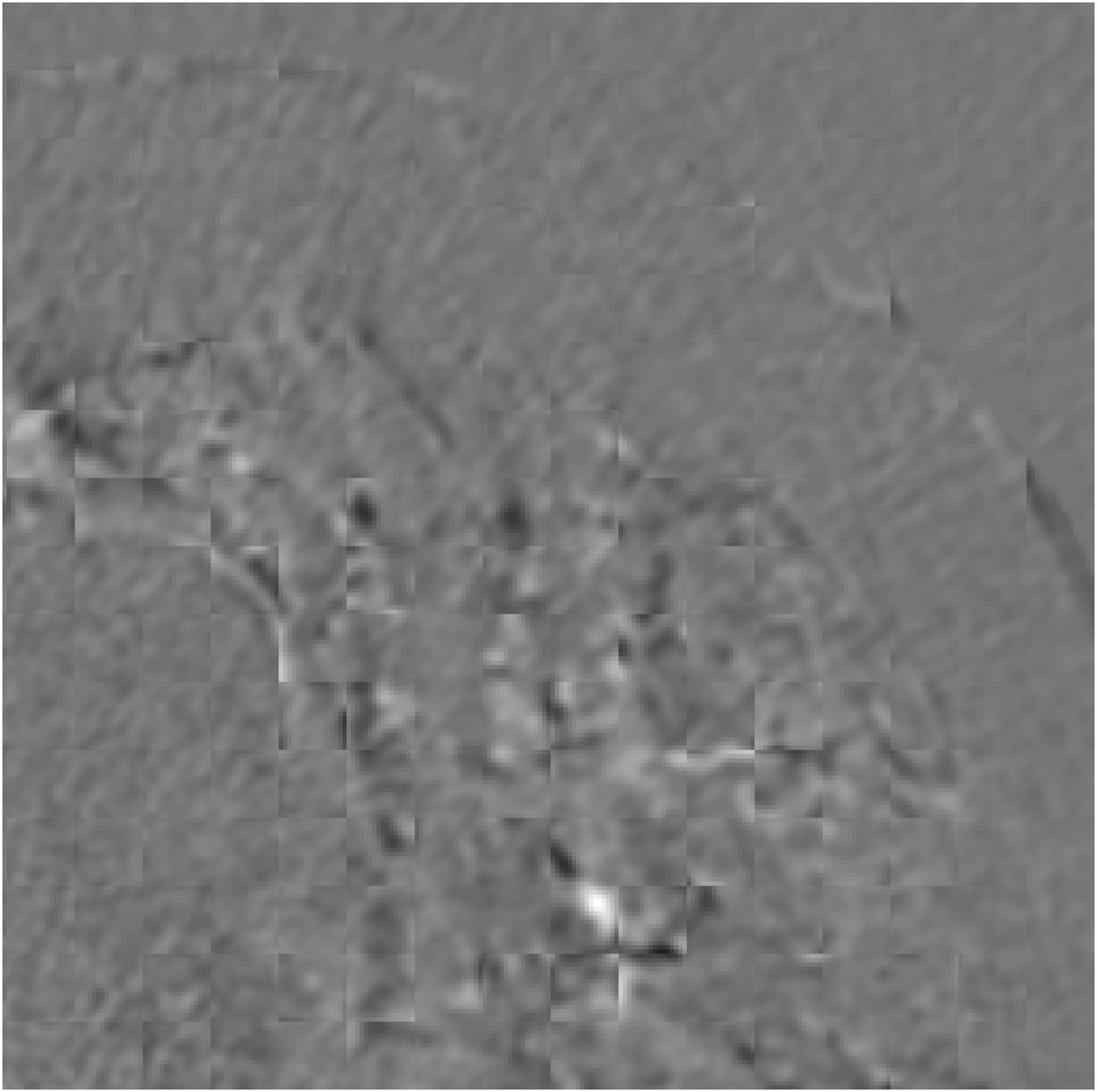}\hfill{}\includegraphics[width=0.32\textwidth]{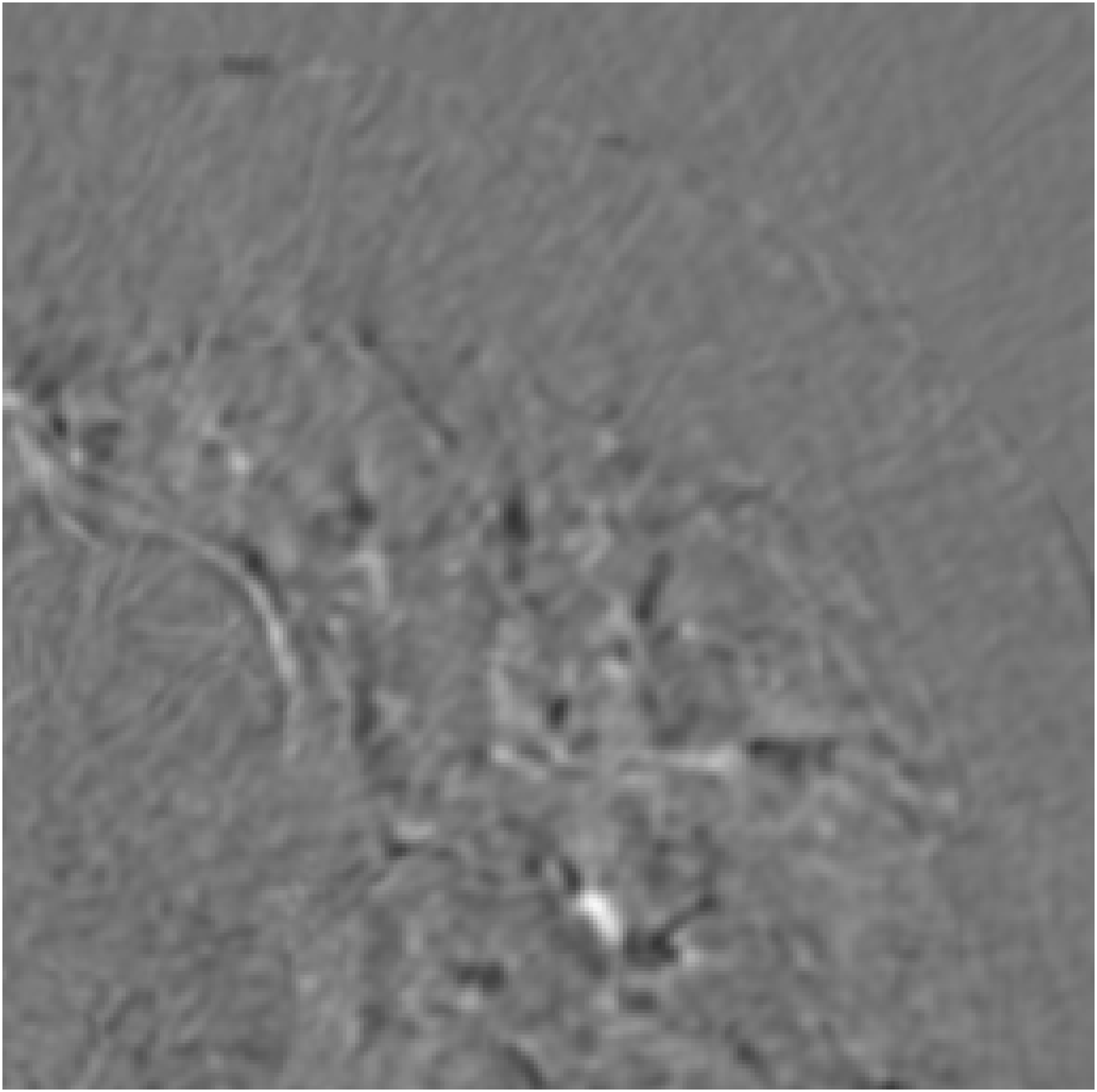}

\hspace{0.1cm}(a) \HP{}-band no compensation\hspace{1.9cm}(b) \HP{}-band
\blockbased{} compensation\hspace{0.6cm}(c) \HP{}-band \meshbased{}
compensation

\vspace{0.3cm}

\includegraphics[width=0.32\textwidth]{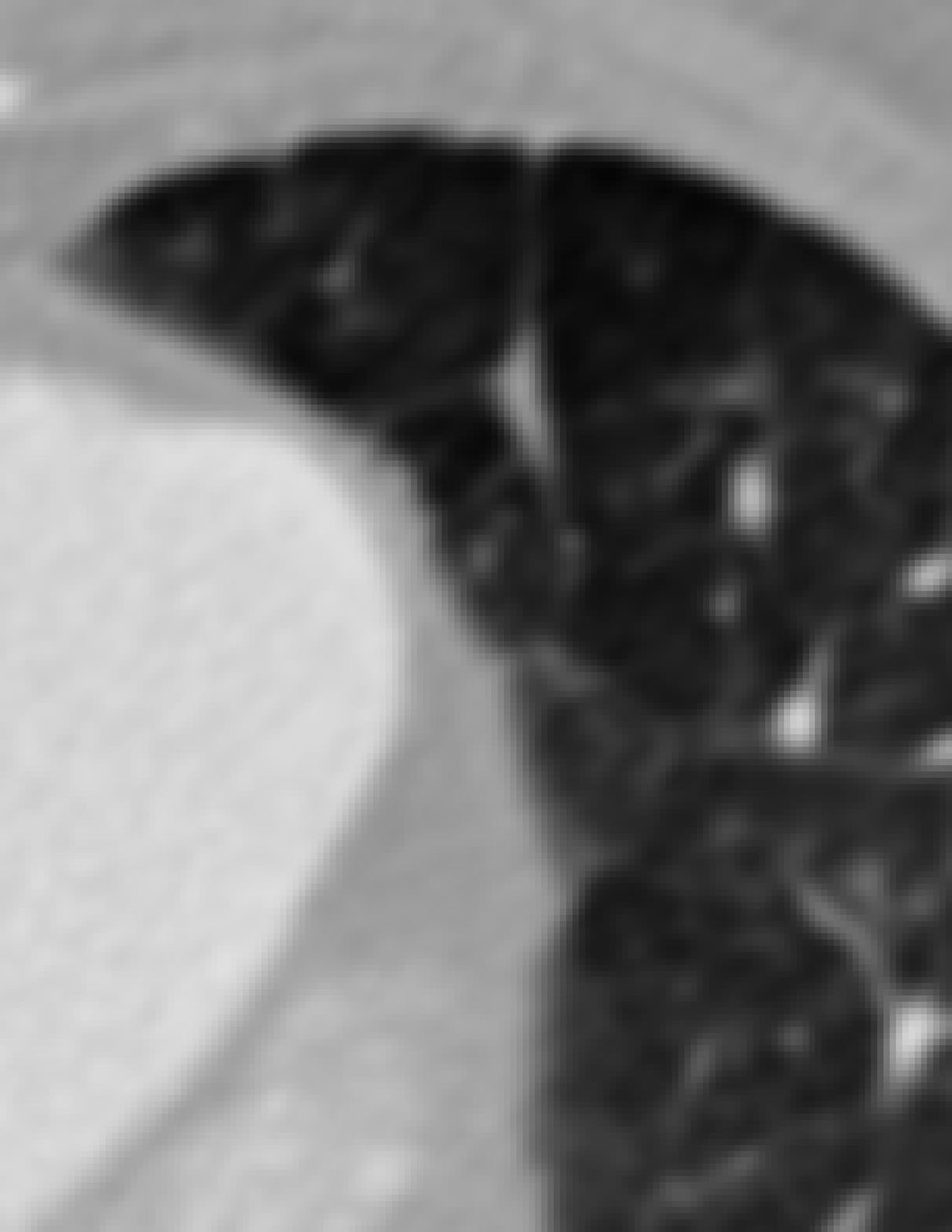}\hfill{}\includegraphics[width=0.32\textwidth]{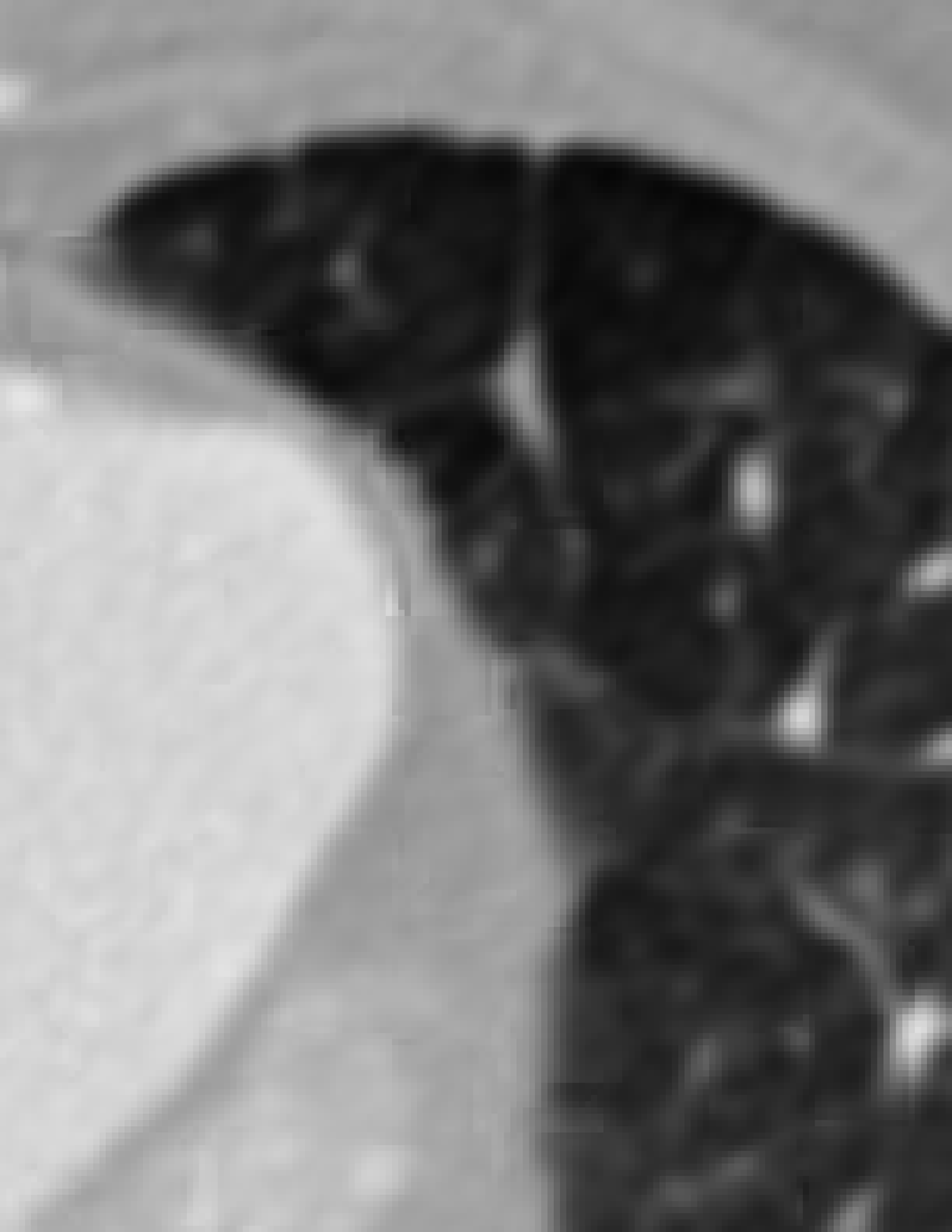}\hfill{}\includegraphics[width=0.32\textwidth]{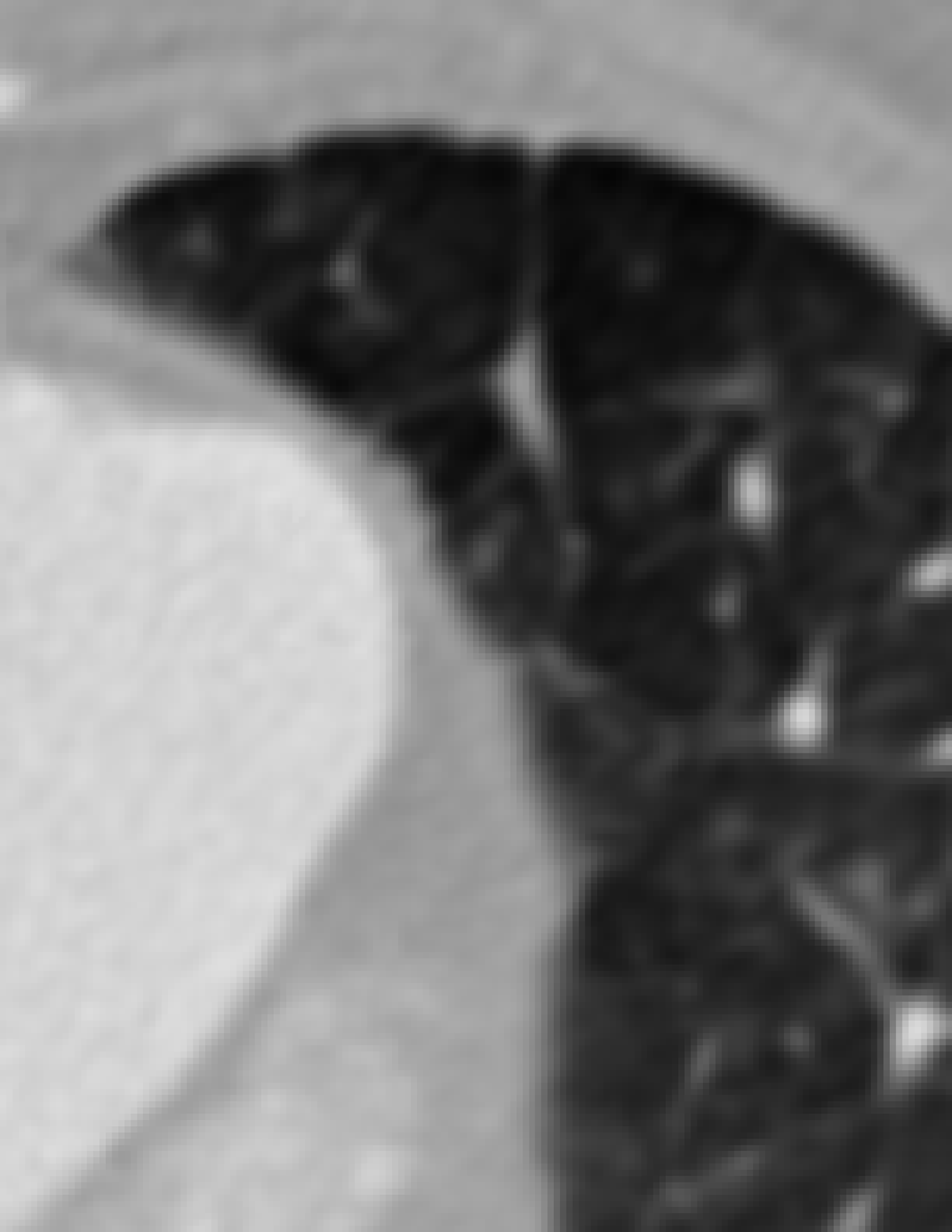}

\hspace{0.1cm}(d) \LP{}-band no compensation\hspace{1.9cm}(e) \LP{}-band
\blockbased{} compensation\hspace{0.6cm}(f) \LP{}-band \meshbased{}
compensation

\protect\caption{\label{fig:Subband-comparison}Zoom into the upper right corner of
the wavelet coefficients: wavelet transform without compensation (a)
and (d) (42.8~dB PSNR of the \LP{}-band compared to the reference
slice), compensated transform with a \blockbased{} method (b) and
(e) (46.1~dB), and compensated transform with a quadrilateral \meshbased{}
method (c) and (f) (46.9~dB), for the upper row: gray is zero, amplitudes
are scaled to the maximal occurring values of the \HP{}-band without
compensation}

\spaceBelowFig
\end{figure*}

For visual comparison, zooms into the upper right quarter of the \HP{}-band
and the \LP{}-band coefficients of the wavelet transform are shown
in \fig{}~\ref{fig:Subband-comparison}. The \HP{}-band coefficients
of the original wavelet transform are shown in \fig{}~\ref{fig:Subband-comparison}~(a).
They contain a lot of energy as can be seen by the distinct edges.
Without compensation, the \LP{}-band, shown in \fig{}~\ref{fig:Subband-comparison}~(d),
gets blurred and contains artifacts which lead to a reduction in visual
quality. The \HP{}-band coefficients of the \blockbased{} method
in \fig{}~\ref{fig:Subband-comparison}~(b) show the typical artifacts
along the block borders. These newly introduced high frequencies have
a very bad impact on the efficiency of a subsequent wavelet-based
coding as discussed later. The advantage of the smooth vector field
from the \meshbased{} method is that there are no blocking artifacts
as can be seen in \fig{}~\ref{fig:Subband-comparison}~(c). Both,
the \blockbased{} and the \meshbased{} method reduce the energy
of the \HP{}-band. The artifacts are hardly visible in the \LP{}-bands
in \fig{}~\ref{fig:Subband-comparison}~(e) and (f) although the
objective quality of the \LP{}-band with \meshbased{} compensation
is better by 0.8~dB PSNR.

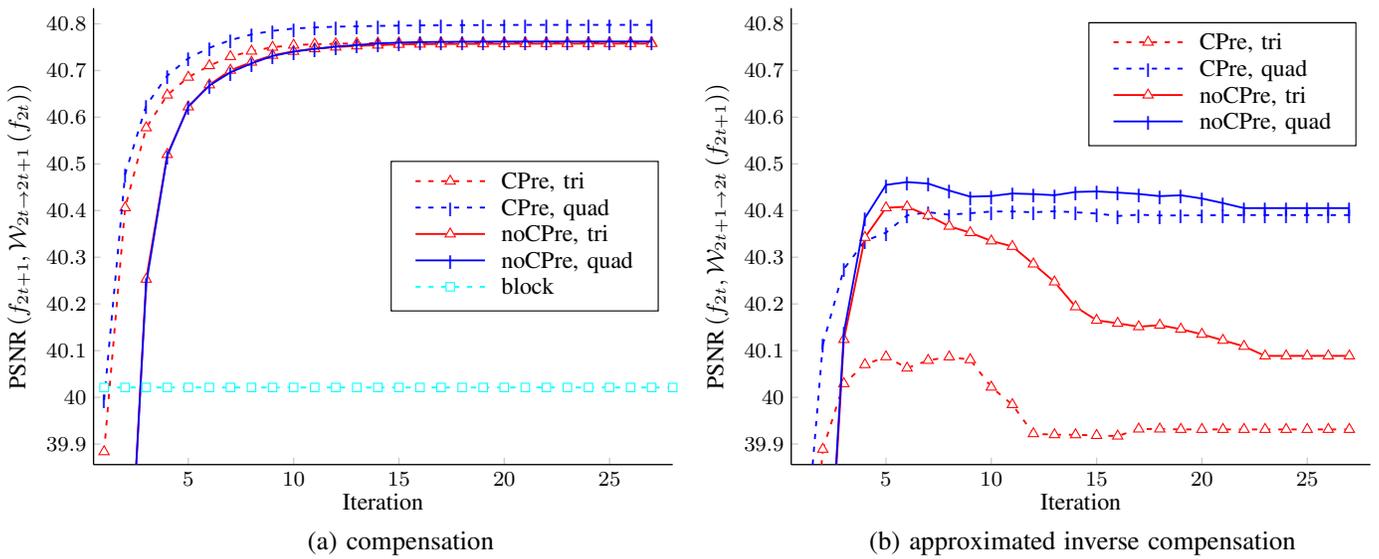
\begin{figure*}
%
%
%
\providelength{\AxesLineWidth}       \setlength{\AxesLineWidth}{0.5pt}%
\providelength{\plotwidth}           \setlength{\plotwidth}{7.7cm}
\providelength{\LineWidth}           \setlength{\LineWidth}{0.7pt}%
\providelength{\MarkerSize}          \setlength{\MarkerSize}{4pt}%
\newrgbcolor{GridColor}{0.8 0.8 0.8}%
%
\psset{xunit=0.036364\plotwidth,yunit=0.806731\plotwidth}%
\begin{pspicture}(-3.607143,39.743259)(28.035714,40.835218)%


\psline[linewidth=\AxesLineWidth,linecolor=GridColor](5.000000,39.855947)(5.000000,39.870822)
\psline[linewidth=\AxesLineWidth,linecolor=GridColor](10.000000,39.855947)(10.000000,39.870822)
\psline[linewidth=\AxesLineWidth,linecolor=GridColor](15.000000,39.855947)(15.000000,39.870822)
\psline[linewidth=\AxesLineWidth,linecolor=GridColor](20.000000,39.855947)(20.000000,39.870822)
\psline[linewidth=\AxesLineWidth,linecolor=GridColor](25.000000,39.855947)(25.000000,39.870822)
\psline[linewidth=\AxesLineWidth,linecolor=GridColor](0.500000,39.900000)(0.830000,39.900000)
\psline[linewidth=\AxesLineWidth,linecolor=GridColor](0.500000,40.000000)(0.830000,40.000000)
\psline[linewidth=\AxesLineWidth,linecolor=GridColor](0.500000,40.100000)(0.830000,40.100000)
\psline[linewidth=\AxesLineWidth,linecolor=GridColor](0.500000,40.200000)(0.830000,40.200000)
\psline[linewidth=\AxesLineWidth,linecolor=GridColor](0.500000,40.300000)(0.830000,40.300000)
\psline[linewidth=\AxesLineWidth,linecolor=GridColor](0.500000,40.400000)(0.830000,40.400000)
\psline[linewidth=\AxesLineWidth,linecolor=GridColor](0.500000,40.500000)(0.830000,40.500000)
\psline[linewidth=\AxesLineWidth,linecolor=GridColor](0.500000,40.600000)(0.830000,40.600000)
\psline[linewidth=\AxesLineWidth,linecolor=GridColor](0.500000,40.700000)(0.830000,40.700000)
\psline[linewidth=\AxesLineWidth,linecolor=GridColor](0.500000,40.800000)(0.830000,40.800000)

{ \footnotesize 
\rput[t](5.000000,39.841072){$5$}
\rput[t](10.000000,39.841072){$10$}
\rput[t](15.000000,39.841072){$15$}
\rput[t](20.000000,39.841072){$20$}
\rput[t](25.000000,39.841072){$25$}
\rput[r](0.170000,39.900000){$39.9$}
\rput[r](0.170000,40.000000){$40$}
\rput[r](0.170000,40.100000){$40.1$}
\rput[r](0.170000,40.200000){$40.2$}
\rput[r](0.170000,40.300000){$40.3$}
\rput[r](0.170000,40.400000){$40.4$}
\rput[r](0.170000,40.500000){$40.5$}
\rput[r](0.170000,40.600000){$40.6$}
\rput[r](0.170000,40.700000){$40.7$}
\rput[r](0.170000,40.800000){$40.8$}
} 

\psline[linewidth=\AxesLineWidth](0.500000,39.855947)(28.000000,39.855947)
\psline[linewidth=\AxesLineWidth](0.500000,39.855947)(0.500000,40.833608)

{ \small 
\rput[b](14.250000,39.743259){
\begin{tabular}{c}
Iteration\\
\end{tabular}
}

\rput[t]{90}(-3.607143,40.344777){
\begin{tabular}{c}
$\text{PSNR} \left( f_{2t+1} , \mathcal{W}_{2t\rightarrow2t+1} \left( \sigf_{2t} \right) \right)$\\
\end{tabular}
}
} 

\newrgbcolor{color477.0352}{1  0  0}
\psline[plotstyle=line,linejoin=1,showpoints=true,dotstyle=Btriangle,dotsize=\MarkerSize,linestyle=dashed,dash=2pt 3pt,linewidth=\LineWidth,linecolor=color477.0352]
(1.000000,39.883470)(2.000000,40.406141)(3.000000,40.577316)(4.000000,40.647406)(5.000000,40.685650)
(6.000000,40.710344)(7.000000,40.729872)(8.000000,40.741500)(9.000000,40.749807)(10.000000,40.754326)
(11.000000,40.756410)(12.000000,40.757245)(13.000000,40.758076)(14.000000,40.758722)(15.000000,40.759052)
(16.000000,40.759222)(17.000000,40.759355)(18.000000,40.759468)(19.000000,40.759538)(20.000000,40.759588)
(21.000000,40.759599)(22.000000,40.759599)(23.000000,40.759599)(24.000000,40.759599)(25.000000,40.759599)
(26.000000,40.759599)(27.000000,40.759599)

\newrgbcolor{color478.0347}{0  0  1}
\psline[plotstyle=line,linejoin=1,showpoints=true,dotstyle=B|,dotsize=\MarkerSize,linestyle=dashed,dash=2pt 3pt,linewidth=\LineWidth,linecolor=color478.0347]
(1.000000,39.994287)(2.000000,40.477874)(3.000000,40.625004)(4.000000,40.688789)(5.000000,40.726193)
(6.000000,40.748627)(7.000000,40.765042)(8.000000,40.777200)(9.000000,40.784952)(10.000000,40.789936)
(11.000000,40.792269)(12.000000,40.793850)(13.000000,40.794692)(14.000000,40.795501)(15.000000,40.796158)
(16.000000,40.796646)(17.000000,40.796987)(18.000000,40.797229)(19.000000,40.797320)(20.000000,40.797526)
(21.000000,40.797686)(22.000000,40.797753)(23.000000,40.797753)(24.000000,40.797753)(25.000000,40.797753)
(26.000000,40.797753)(27.000000,40.797753)

\newrgbcolor{color479.0347}{1  0  0}
\psline[plotstyle=line,linejoin=1,showpoints=false,dotstyle=Btriangle,dotsize=\MarkerSize,linestyle=solid,linewidth=\LineWidth,linecolor=color479.0347]
(2.525834,39.855947)(3.000000,40.252988)
\psline[plotstyle=line,linejoin=1,showpoints=true,dotstyle=Btriangle,dotsize=\MarkerSize,linestyle=solid,linewidth=\LineWidth,linecolor=color479.0347]
(3.000000,40.252988)(4.000000,40.520049)(5.000000,40.622051)(6.000000,40.669268)(7.000000,40.700544)
(8.000000,40.717883)(9.000000,40.732593)(10.000000,40.740980)(11.000000,40.746590)(12.000000,40.750197)
(13.000000,40.752680)(14.000000,40.754368)(15.000000,40.755971)(16.000000,40.756460)(17.000000,40.756840)
(18.000000,40.757061)(19.000000,40.757171)(20.000000,40.757289)(21.000000,40.757483)(22.000000,40.757558)
(23.000000,40.757586)(24.000000,40.757680)(25.000000,40.757684)(26.000000,40.757684)(27.000000,40.757684)

\newrgbcolor{color480.0347}{0  0  1}
\psline[plotstyle=line,linejoin=1,showpoints=false,dotstyle=B|,dotsize=\MarkerSize,linestyle=solid,linewidth=\LineWidth,linecolor=color480.0347]
(2.528235,39.855947)(3.000000,40.243163)
\psline[plotstyle=line,linejoin=1,showpoints=true,dotstyle=B|,dotsize=\MarkerSize,linestyle=solid,linewidth=\LineWidth,linecolor=color480.0347]
(3.000000,40.243163)(4.000000,40.517200)(5.000000,40.622919)(6.000000,40.667747)(7.000000,40.696051)
(8.000000,40.715203)(9.000000,40.731109)(10.000000,40.740431)(11.000000,40.746380)(12.000000,40.751089)
(13.000000,40.754556)(14.000000,40.757939)(15.000000,40.759816)(16.000000,40.760917)(17.000000,40.761258)
(18.000000,40.761605)(19.000000,40.761825)(20.000000,40.761950)(21.000000,40.762051)(22.000000,40.762117)
(23.000000,40.762117)(24.000000,40.762117)(25.000000,40.762117)(26.000000,40.762117)(27.000000,40.762117)

\newrgbcolor{color481.0347}{0  1  1}
\psline[plotstyle=line,linejoin=1,showpoints=false,dotstyle=Bsquare,dotsize=\MarkerSize,linestyle=dashed,dash=2pt 3pt,linewidth=\LineWidth,linecolor=color481.0347]
(28.000000,40.021100)(28.000000,40.021100)
\psline[plotstyle=line,linejoin=1,showpoints=true,dotstyle=Bsquare,dotsize=\MarkerSize,linestyle=dashed,dash=2pt 3pt,linewidth=\LineWidth,linecolor=color481.0347]
(1.000000,40.021100)(2.000000,40.021100)(3.000000,40.021100)(4.000000,40.021100)(5.000000,40.021100)
(6.000000,40.021100)(7.000000,40.021100)(8.000000,40.021100)(9.000000,40.021100)(10.000000,40.021100)
(11.000000,40.021100)(12.000000,40.021100)(13.000000,40.021100)(14.000000,40.021100)(15.000000,40.021100)
(16.000000,40.021100)(17.000000,40.021100)(18.000000,40.021100)(19.000000,40.021100)(20.000000,40.021100)
(21.000000,40.021100)(22.000000,40.021100)(23.000000,40.021100)(24.000000,40.021100)(25.000000,40.021100)
(26.000000,40.021100)(27.000000,40.021100)(28.000000,40.021100)

{ \small 
\rput[r](27.340000,40.344777){%
\psframebox[framesep=0pt,linewidth=\AxesLineWidth]{\psframebox*{\begin{tabular}{l}
\Rnode{a1}{\hspace*{0.0ex}} \hspace*{0.7cm} \Rnode{a2}{~~CPre, tri} \\
\Rnode{a3}{\hspace*{0.0ex}} \hspace*{0.7cm} \Rnode{a4}{~~CPre, quad} \\
\Rnode{a5}{\hspace*{0.0ex}} \hspace*{0.7cm} \Rnode{a6}{~~noCPre, tri} \\
\Rnode{a7}{\hspace*{0.0ex}} \hspace*{0.7cm} \Rnode{a8}{~~noCPre, quad} \\
\Rnode{a9}{\hspace*{0.0ex}} \hspace*{0.7cm} \Rnode{a10}{~~block} \\
\end{tabular}}
\ncline[linestyle=dashed,dash=2pt 3pt,linewidth=\LineWidth,linecolor=color477.0352]{a1}{a2} \ncput{\psdot[dotstyle=Btriangle,dotsize=\MarkerSize,linecolor=color477.0352]}
\ncline[linestyle=dashed,dash=2pt 3pt,linewidth=\LineWidth,linecolor=color478.0347]{a3}{a4} \ncput{\psdot[dotstyle=B|,dotsize=\MarkerSize,linecolor=color478.0347]}
\ncline[linestyle=solid,linewidth=\LineWidth,linecolor=color479.0347]{a5}{a6} \ncput{\psdot[dotstyle=Btriangle,dotsize=\MarkerSize,linecolor=color479.0347]}
\ncline[linestyle=solid,linewidth=\LineWidth,linecolor=color480.0347]{a7}{a8} \ncput{\psdot[dotstyle=B|,dotsize=\MarkerSize,linecolor=color480.0347]}
\ncline[linestyle=dashed,dash=2pt 3pt,linewidth=\LineWidth,linecolor=color481.0347]{a9}{a10} \ncput{\psdot[dotstyle=Bsquare,dotsize=\MarkerSize,linecolor=color481.0347]}
}%
}%
} 

\end{pspicture}%
\hfill{}
%
%
%
\providelength{\AxesLineWidth}       \setlength{\AxesLineWidth}{0.5pt}%
\providelength{\plotwidth}           \setlength{\plotwidth}{7.7cm}
\providelength{\LineWidth}           \setlength{\LineWidth}{0.7pt}%
\providelength{\MarkerSize}          \setlength{\MarkerSize}{4pt}%
\newrgbcolor{GridColor}{0.8 0.8 0.8}%
%
\psset{xunit=0.036364\plotwidth,yunit=0.806731\plotwidth}%
\begin{pspicture}(-3.607143,39.743259)(28.035714,40.835218)%


\psline[linewidth=\AxesLineWidth,linecolor=GridColor](5.000000,39.855947)(5.000000,39.870822)
\psline[linewidth=\AxesLineWidth,linecolor=GridColor](10.000000,39.855947)(10.000000,39.870822)
\psline[linewidth=\AxesLineWidth,linecolor=GridColor](15.000000,39.855947)(15.000000,39.870822)
\psline[linewidth=\AxesLineWidth,linecolor=GridColor](20.000000,39.855947)(20.000000,39.870822)
\psline[linewidth=\AxesLineWidth,linecolor=GridColor](25.000000,39.855947)(25.000000,39.870822)
\psline[linewidth=\AxesLineWidth,linecolor=GridColor](0.500000,39.900000)(0.830000,39.900000)
\psline[linewidth=\AxesLineWidth,linecolor=GridColor](0.500000,40.000000)(0.830000,40.000000)
\psline[linewidth=\AxesLineWidth,linecolor=GridColor](0.500000,40.100000)(0.830000,40.100000)
\psline[linewidth=\AxesLineWidth,linecolor=GridColor](0.500000,40.200000)(0.830000,40.200000)
\psline[linewidth=\AxesLineWidth,linecolor=GridColor](0.500000,40.300000)(0.830000,40.300000)
\psline[linewidth=\AxesLineWidth,linecolor=GridColor](0.500000,40.400000)(0.830000,40.400000)
\psline[linewidth=\AxesLineWidth,linecolor=GridColor](0.500000,40.500000)(0.830000,40.500000)
\psline[linewidth=\AxesLineWidth,linecolor=GridColor](0.500000,40.600000)(0.830000,40.600000)
\psline[linewidth=\AxesLineWidth,linecolor=GridColor](0.500000,40.700000)(0.830000,40.700000)
\psline[linewidth=\AxesLineWidth,linecolor=GridColor](0.500000,40.800000)(0.830000,40.800000)

{ \footnotesize 
\rput[t](5.000000,39.841072){$5$}
\rput[t](10.000000,39.841072){$10$}
\rput[t](15.000000,39.841072){$15$}
\rput[t](20.000000,39.841072){$20$}
\rput[t](25.000000,39.841072){$25$}
\rput[r](0.170000,39.900000){$39.9$}
\rput[r](0.170000,40.000000){$40$}
\rput[r](0.170000,40.100000){$40.1$}
\rput[r](0.170000,40.200000){$40.2$}
\rput[r](0.170000,40.300000){$40.3$}
\rput[r](0.170000,40.400000){$40.4$}
\rput[r](0.170000,40.500000){$40.5$}
\rput[r](0.170000,40.600000){$40.6$}
\rput[r](0.170000,40.700000){$40.7$}
\rput[r](0.170000,40.800000){$40.8$}
} 

\psline[linewidth=\AxesLineWidth](0.500000,39.855947)(28.000000,39.855947)
\psline[linewidth=\AxesLineWidth](0.500000,39.855947)(0.500000,40.833608)

{ \small 
\rput[b](14.250000,39.743259){
\begin{tabular}{c}
Iteration\\
\end{tabular}
}

\rput[t]{90}(-3.607143,40.344777){
\begin{tabular}{c}
$\text{PSNR} \left( f_{2t} , \mathcal{W}_{2t+1\rightarrow2t} \left( \sigf_{2t+1} \right) \right)$\\
\end{tabular}
}
} 

\newrgbcolor{color473.0353}{1  0  0}
\psline[plotstyle=line,linejoin=1,showpoints=false,dotstyle=Btriangle,dotsize=\MarkerSize,linestyle=dashed,dash=2pt 3pt,linewidth=\LineWidth,linecolor=color473.0353]
(1.939487,39.855947)(2.000000,39.888754)
\psline[plotstyle=line,linejoin=1,showpoints=true,dotstyle=Btriangle,dotsize=\MarkerSize,linestyle=dashed,dash=2pt 3pt,linewidth=\LineWidth,linecolor=color473.0353]
(2.000000,39.888754)(3.000000,40.029222)(4.000000,40.070032)(5.000000,40.086915)(6.000000,40.062655)
(7.000000,40.079182)(8.000000,40.086564)(9.000000,40.080595)(10.000000,40.021837)(11.000000,39.984347)
(12.000000,39.922129)(13.000000,39.919916)(14.000000,39.919834)(15.000000,39.918327)(16.000000,39.917037)
(17.000000,39.932127)(18.000000,39.932147)(19.000000,39.931299)(20.000000,39.931263)(21.000000,39.931290)
(22.000000,39.931290)(23.000000,39.931290)(24.000000,39.931290)(25.000000,39.931290)(26.000000,39.931290)
(27.000000,39.931290)

\newrgbcolor{color474.0348}{0  0  1}
\psline[plotstyle=line,linejoin=1,showpoints=false,dotstyle=B|,dotsize=\MarkerSize,linestyle=dashed,dash=2pt 3pt,linewidth=\LineWidth,linecolor=color474.0348]
(1.517011,39.855947)(2.000000,40.113290)
\psline[plotstyle=line,linejoin=1,showpoints=true,dotstyle=B|,dotsize=\MarkerSize,linestyle=dashed,dash=2pt 3pt,linewidth=\LineWidth,linecolor=color474.0348]
(2.000000,40.113290)(3.000000,40.274909)(4.000000,40.334856)(5.000000,40.352732)(6.000000,40.389717)
(7.000000,40.396510)(8.000000,40.391178)(9.000000,40.394410)(10.000000,40.397945)(11.000000,40.398185)
(12.000000,40.396363)(13.000000,40.398627)(14.000000,40.396356)(15.000000,40.393704)(16.000000,40.388375)
(17.000000,40.391141)(18.000000,40.389574)(19.000000,40.389633)(20.000000,40.389890)(21.000000,40.390190)
(22.000000,40.390420)(23.000000,40.390420)(24.000000,40.390420)(25.000000,40.390420)(26.000000,40.390420)
(27.000000,40.390420)

\newrgbcolor{color475.0348}{1  0  0}
\psline[plotstyle=line,linejoin=1,showpoints=false,dotstyle=Btriangle,dotsize=\MarkerSize,linestyle=solid,linewidth=\LineWidth,linecolor=color475.0348]
(2.651429,39.855947)(3.000000,40.123856)
\psline[plotstyle=line,linejoin=1,showpoints=true,dotstyle=Btriangle,dotsize=\MarkerSize,linestyle=solid,linewidth=\LineWidth,linecolor=color475.0348]
(3.000000,40.123856)(4.000000,40.342504)(5.000000,40.406066)(6.000000,40.408289)(7.000000,40.389347)
(8.000000,40.366423)(9.000000,40.352716)(10.000000,40.334981)(11.000000,40.323242)(12.000000,40.285648)
(13.000000,40.247151)(14.000000,40.193919)(15.000000,40.165007)(16.000000,40.158623)(17.000000,40.151240)
(18.000000,40.154631)(19.000000,40.145961)(20.000000,40.134982)(21.000000,40.122138)(22.000000,40.109250)
(23.000000,40.088679)(24.000000,40.088821)(25.000000,40.088785)(26.000000,40.088785)(27.000000,40.088785)

\newrgbcolor{color476.0352}{0  0  1}
\psline[plotstyle=line,linejoin=1,showpoints=false,dotstyle=B|,dotsize=\MarkerSize,linestyle=solid,linewidth=\LineWidth,linecolor=color476.0352]
(2.628139,39.855947)(3.000000,40.138751)
\psline[plotstyle=line,linejoin=1,showpoints=true,dotstyle=B|,dotsize=\MarkerSize,linestyle=solid,linewidth=\LineWidth,linecolor=color476.0352]
(3.000000,40.138751)(4.000000,40.384081)(5.000000,40.455018)(6.000000,40.461056)(7.000000,40.457875)
(8.000000,40.442905)(9.000000,40.429880)(10.000000,40.431097)(11.000000,40.436624)(12.000000,40.435273)
(13.000000,40.432902)(14.000000,40.439881)(15.000000,40.440997)(16.000000,40.438626)(17.000000,40.435495)
(18.000000,40.430940)(19.000000,40.432774)(20.000000,40.425871)(21.000000,40.416029)(22.000000,40.405318)
(23.000000,40.405318)(24.000000,40.405318)(25.000000,40.405318)(26.000000,40.405318)(27.000000,40.405318)

{ \small 
\rput[tr](27.340000,40.803858){%
\psframebox[framesep=0pt,linewidth=\AxesLineWidth]{\psframebox*{\begin{tabular}{l}
\Rnode{a1}{\hspace*{0.0ex}} \hspace*{0.7cm} \Rnode{a2}{~~CPre, tri} \\
\Rnode{a3}{\hspace*{0.0ex}} \hspace*{0.7cm} \Rnode{a4}{~~CPre, quad} \\
\Rnode{a5}{\hspace*{0.0ex}} \hspace*{0.7cm} \Rnode{a6}{~~noCPre, tri} \\
\Rnode{a7}{\hspace*{0.0ex}} \hspace*{0.7cm} \Rnode{a8}{~~noCPre, quad} \\
\end{tabular}}
\ncline[linestyle=dashed,dash=2pt 3pt,linewidth=\LineWidth,linecolor=color473.0353]{a1}{a2} \ncput{\psdot[dotstyle=Btriangle,dotsize=\MarkerSize,linecolor=color473.0353]}
\ncline[linestyle=dashed,dash=2pt 3pt,linewidth=\LineWidth,linecolor=color474.0348]{a3}{a4} \ncput{\psdot[dotstyle=B|,dotsize=\MarkerSize,linecolor=color474.0348]}
\ncline[linestyle=solid,linewidth=\LineWidth,linecolor=color475.0348]{a5}{a6} \ncput{\psdot[dotstyle=Btriangle,dotsize=\MarkerSize,linecolor=color475.0348]}
\ncline[linestyle=solid,linewidth=\LineWidth,linecolor=color476.0352]{a7}{a8} \ncput{\psdot[dotstyle=B|,dotsize=\MarkerSize,linecolor=color476.0352]}
}%
}%
} 

\end{pspicture}%

\hspace{4cm}(a) compensation\hspace{5cm}(b) approximated inverse
compensation

\protect\caption{\label{fig:convergence-speed}Convergence speed and performance of
the considered compensation methods, left: compensation, right: approximated
inverse compensation}
\spaceBelowFig
\end{figure*}

The two plots in \fig{}~\ref{fig:convergence-speed} show the convergence
speed in addition to the performance of the considered compensation
methods in terms of PSNR. For the case that all information is used
for decoding, the lifting structure guarantees perfect reconstruction.
We calculate the PSNR to evaluate the scalability, i.e., for the case
that only the \LP{}-band is decoded as scaled version according to
the temporal axis. The higher the PSNR the less artifacts are contained
in the \LP{}-band. \fig{}~\ref{fig:convergence-speed}~(a) shows
the results for the compensation for the triangle mesh `tri' and the
quadrilateral mesh, denoted by `quad'.  `CPre' indicates that a \blockbased{}
coarse estimation was performed while `noCPre' indicates that the
refinement was initialized with a zero motion vector field. 

For comparison, the PSNR of the \blockbased{} compensation is also
plotted as straight line as it does not depend on the number of the
refinement iteration. After the third iteration, the \meshbased{}
approaches reach a higher PSNR then the \blockbased{} method. The
plot shows that by an iteration number of 15 the convergence is basically
reached. This number of iterations is also recommended as the PSNR
is improved significantly until then. The plot shows that the mesh-based
methods achieve comparable results for the compensation step. The
performance of the coarse estimation depends on the image content
but the quadrilateral mesh achieved the best results in our simulations.
The reason for the better performance of the quadrilateral mesh is
the smoother vector field due to the bilinear transformation.

\fig{}~\ref{fig:convergence-speed}~(b) shows the results for the
approximation of the inverse compensation when the mesh of the corresponding
iteration is used. The highpass frame contains only noisy structures
so a direct evaluation of the quality of the inverse warping is difficult.
To be able to evaluate the quality of the inversion we compute the
difference between the inverse warped current slice and the reference
slice instead.  The curves for the inversion drop with increasing
iteration number. The reason is that the optimization is performed
for the compensation step. In every iteration, the motion vectors
can become larger and that can enlarge the error of the approximation
of the inversion. Nevertheless the PSNR values of the approximation
of the inverse compensation are in a similar range compared to the
PSNR values of the compensation in \fig{}~\ref{fig:convergence-speed}~(a).
The triangle mesh is more susceptible to the incorporated approximation.
The quadrilateral mesh achieves better results for the approximated
inverse compensation.

To evaluate the compressibility of the different approaches we coded
the resulting wavelet coefficients with JPEG~2000 \cite{ITU-T-800},
where a further wavelet decomposition of 4 steps in $x$-$y$ direction
is performed. The results of the 10 temporally adjacent slices are
listed in \tab{}~\ref{tab:Coding-results}. The Haar wavelet processes
two slices at a time, so all bits needed for each two slices are summed
up as listed in the second row. For comparison, the result of the
 coded original slices $f_{2i}$ and $f_{2i+1}$ is given in the
second column.  The amount of side information for the motion vectors
$\text{MV}_{t}$, which is in the order of about 1~kbyte, is similar
for the \blockbased{} approach and the \meshbased{} approach. The
transform with \blockbased{} compensation needs the most bits for
coding. The results for the \meshbased{} compensation method with
quadrilateral topology show that there are gains possible against
the common slice-wise coding of the original slices. The reason for
the good compressibility of the uncompensated transform is the amount
of correlated noisy structures that are compensated in this case
as neighboring pixels are processed together. A compensation method
leads to processing of neighboring pixels according to the structural
information. But the uncompensated transform has the disadvantage
that the blur of the \LP{}-band leads to a decrease of quality as
the PSNR values show in \tab{}~\ref{tab:scale-results}. The blur
increases with every further decomposition step which leads to a poor
scalable representation. This can be prevented by a compensation method
in the lifting structure as  shown for the \blockbased{} and the
\meshbased{} method. The \meshbased{} method achieves a higher PSNR
and a smaller filesize compared to the \blockbased{} method.

\begin{table}
\protect\caption{\label{tab:Coding-results}JPEG~2000 lossless coding results: Filesize
in kbyte for \textit{cardiac}}
\vspace{-2mm}

\begin{tabular}{|c|c|c|c|c|}
\hline
$t$ & slice-wise & no  & \blockbased{}  & \meshbased{}  \tabularnewline
 &  &  compensation &  compensation &  compensation \tabularnewline
\hline 
 & $f_{2t}$, $f_{2t+1}$ & $\HP_{t}$, $\LP_t$ & $\HP_{t}$, $\LP_t$, $\text{MV}_t$ & $\HP_{t}$, $\LP_t$, $\text{MV}_t$ \tabularnewline
\hline 
\hline
1 & 295.8 & 286.9 & 307.4 & 294 \tabularnewline 

2 & 297.9 & 283.2 & 303.2 & 290.8 \tabularnewline 

3 & 298.7 & 283.2 & 309.2 & 296.2 \tabularnewline 

4 & 297.9 & 279.6 & 299.5 & 288.7 \tabularnewline 

5 & 296.8 & 283.1 & 299.5 & 288.8 \tabularnewline 
\tabularnewline\hline
\end{tabular}
\end{table}
\begin{table}
\protect\caption{\label{tab:scale-results}\LP{}-band results: JPEG~2000 filesize
in kbyte, $\text{PSNR}\left(\LP_{t},f_{2t}\right)$, averaged over
all time steps $t$}
\vspace{-2mm}

\begin{tabular*}{0.98\columnwidth}{@{\extracolsep{\fill}}|c|c|c|c|c|c|} 
\hline  
\multicolumn{2}{|c|}{no} & \multicolumn{2}{c|}{\blockbased{} } & \multicolumn{2}{c|}{\meshbased{} }\tabularnewline
\multicolumn{2}{|c|}{compensation} & \multicolumn{2}{c|}{ compensation} & \multicolumn{2}{c|}{ compensation}\tabularnewline
\hline  
\hline
 filesize & PSNR & filesize & PSNR & filesize & PSNR \tabularnewline
\hline

 142.7 & 43.5 dB & 151.3 & 46.3 dB & 148.1 & 47.3 dB \tabularnewline

\hline
\end{tabular*}

\spaceBelowTab
\end{table}

As long as no inversion of the mesh is needed, the grid size can be
chosen small. But the error of the approximation of the inverse compensation
grows with a smaller block size. For compensation in lifted wavelet-based
approaches, where the inversion is necessary in the update step, either
the block size has to be chosen big enough or a refinement mesh is
needed for the update step. While the results of the triangle-based
mesh are comparable to the results of the quadrilateral-based mesh
in the compensation step, the quadrilateral-based mesh yields better
results in the inverse compensation step.

For a mesh-based compensation method of dynamic cardiac CT data in
temporal direction, the \blockbased{} estimation of the movement
of the grid points can be used as a starting point for an iterative
refinement. Our results show that about 15 refinement iteration steps
are necessary for convergence. Then the results are very similar
whether a coarse estimation is used or not. Actually, in some cases
there were better results in the motion compensation step without
the \blockbased{} coarse estimation. In the inversion step the results
of the approximation of the method without coarse estimation works
better in near all cases.

\section{Conclusion}

A lifted wavelet transform can be adapted to the signal by extending
the transform with a proper compensation method. When the compensation
method is able to compensate the displacement in transform direction
the otherwise occurring artifacts in the \LP-band{} can be reduced.
In this paper we showed that \meshbased{} methods are feasible for
the compensation of the displacement in the temporal direction of
dynamical cardiac CT volume data. By reducing the artifacts, the quality
of the \LP-band{} is improved. This makes the \LP{}-band more usable
as scalable representation of the volume in the temporal direction.
A scalable representation is very advantageous, e.g., for telemedical
applications. We further could show that for dynamical cardiac CT
volume data \meshbased{} compensation methods are superior to \blockbased{}
compensation for coding of the subbands with JPEG~2000.

Further work aims at a more detailed and theoretic analysis of the
occurring approximation error for the inversion of the mesh. Further
the extension of the \meshbased{} approach to three dimensions will
be considered.

\section*{Acknowledgment}

We gratefully acknowledge that this work has been supported by the
Deutsche Forschungsgemeinschaft (DFG) under contract number KA~926/4-1.

\end{document}